\documentclass[journal=jacsat,manuscript=article]{achemso}
\setkeys{acs}{
  maxauthors = 10,      
  etalmode = truncate   
}

\usepackage[version=3]{mhchem} 
\SectionNumbersOn
\usepackage[table,xcdraw,dvipsnames]{xcolor}

\usepackage{ulem}
\usepackage{booktabs}
\usepackage{hyperref}
\usepackage{cleveref}

\newcommand{\onlinecite}[1]{\hspace{-1 ex} \nocite{#1}\citenum{#1}}

\newcommand{\bochum}{Research Center Future Energy Materials and Systems of the University Alliance Ruhr and Interdisciplinary Centre for Advanced Materials Simulation, Faculty of Physics and Astronomy, Ruhr University Bochum, Universitätsstraße 150, D-44801 Bochum, Germany}
\newcommand{\etsf}{European Theoretical Spectroscopy Facility (ETSF)}

\author{Ayoub Aouina }
\affiliation{\bochum}
\alsoaffiliation{\etsf}

\author{ Nicolas Tancogne-Dejean }
\affiliation{\bochum}
\alsoaffiliation{\etsf}

\author{Silvana Botti}
\affiliation{\bochum}
\alsoaffiliation{\etsf} 
\email{silvana.botti@rub.de}

\title{Functional and Density-Driven Errors in Density Functional Theory: Quantum Monte Carlo Benchmarks for Solids}

\begin{document}


\begin{abstract}
We introduce a systematic analysis of density functional approximation errors in solids by separating functional-driven from density-driven contributions using quantum Monte Carlo densities of silicon, sodium chloride, and copper as reference. Typically, functional errors dominate, but we identify important exceptions where density-driven errors exceed functional errors by factors of 2--3, notably for SOGGA11 and $\tau$-HCTH in the semiconductor and the insulator. 
Material dependence is striking: 63\% of functionals show error cancellation in silicon versus 18\% in copper, and only five functionals surpass LDA accuracy for metallic copper even with exact densities. For silicon and sodium chloride, GILL or BECKE exchange combined with PBE, PW91, or P86 correlation achieves near-exact xc energies on QMC densities, while copper requires specialized functionals like PBEsol or PBELYP. High-quality densities consistently reduce density-driven errors across all systems. Historical analysis reveals that 1990s GGA functionals outperform many modern meta-GGAs, contradicting expectations of systematic improvement along Jacob's ladder. These results provide practical guidance for functional selection and highlight implications for machine learning potential development, where material-dependent error cancellation may compromise transferability. 
\end{abstract}

\section{Introduction}
Density Functional Theory~\cite{HohenbergKohn1964} (DFT) is the most widely used electronic structure method in condensed matter physics and quantum chemistry, offering a balance between computational efficiency and accuracy. Its success relies on the Kohn-Sham (KS) formalism~\cite{KohnSham1965}, which reformulates the many-electron problem into an effective single-particle framework. The key approximation in DFT lies in the exchange-correlation (xc) functional, which must be approximated in practical calculations. The first such approximation, the Local Density Approximation (LDA), was introduced in the same seminal paper as the KS equations~\cite{KohnSham1965} and was motivated by the nearsightedness of electrons in condensed matter systems~\cite{prodan2005}. Since then, hundreds of density functional approximations (DFAs) have been developed for the xc functional~\cite{marques2012,lehtola2018}.

Perdew's ``Jacob's ladder'' metaphor suggests that systematic improvement can be achieved by ascending from LDA to generalized gradient approximation (GGA), meta-GGA, and beyond, incorporating increasingly sophisticated ingredients such as density gradients, kinetic energy density, and exact exchange. However, this appealing picture does not always hold in practice. Our recent comprehensive assessment of xc functionals revealed that GGA functionals from the 1990s often outperform more recent, higher-rung approximations for both the xc energy and electron density of solids~\cite{Aouina2024}. This finding suggests that climbing Jacob's ladder does not guarantee monotonic improvement across all material classes, and raises fundamental questions about the nature of errors in DFT calculations.

Across all functional types, errors remain originating not only from the approximate form of the functional itself but also from inaccuracies in the resulting electron density~\cite{medvedev2017}. Understanding and quantifying these two sources of error is crucial for advancing functional development and improving predictive accuracy. To address this challenge, \citet{kim2013} introduced a powerful framework for decomposing the total energy error of any DFT functional into a \textit{functional error} (reflecting the quality of the approximate xc functional evaluated on the exact density) and a \textit{density-driven error} (arising from using an approximate density). While their analysis focused primarily on atoms and molecules, extended systems remained largely unexplored. Here, we apply this framework systematically to solids, where the interplay between functional approximation and density errors may differ fundamentally from molecular systems due to extended bonding, metallic behavior, and reduced density fluctuations.

The decomposition proceeds as follows. In ground-state DFT, the total energy functional is given by:  
\begin{equation}
\label{eq:energy-functional}
    E[n]= T_s[n] + E_{\rm Hartree}[n] + E_{\rm xc}[n]   + \int d{\bf r} n({\bf r}) v({\bf r}),
\end{equation}  
where $T_s[n]$ is the Kohn-Sham (non-interacting) kinetic energy, $E_{\rm Hartree}[n]$ is the Hartree energy, $E_{\rm xc}[n]$ is the xc functional, the only unknown component requiring approximation, and $v({\bf r})$ is the one-body potential (typically the external potential from nuclei or ions). The exact ground-state total energy is obtained by minimizing Eq.~\eqref{eq:energy-functional} with respect to the density, yielding $E[n^{\rm ref}]$, where $n^{\rm ref}({\bf r})$ is the exact ground-state electron density. 

In practical DFT calculations, $E_{\rm xc}[n]$ is approximated as $E_{\rm xc}^{\rm DFA}[n]$. Solving the self-consistent KS equations with this approximation minimizes the corresponding approximate functional $E^{\rm DFA}[n]$, yielding an approximate density $n^{\rm DFA}({\bf r})$. Crucially, while $E[n^{\rm ref}]$ represents the energy of the exact functional evaluated on the exact density, $E^{\rm DFA}[n^{\rm ref}]$ represents the approximate functional evaluated on the exact density. This distinction allows us to isolate the intrinsic quality of the functional from errors induced by using an approximate density.

The total energy error can then be rigorously decomposed into functional and density-driven components:
\begin{equation}
    \Delta E = E^{\rm DFA}[n^{\rm DFA}] - E[n^{\rm ref}] = \Delta E^F_{\rm xc} + \Delta E^D.
\end{equation} 
The functional error, $\Delta E^F_{\rm xc}$, quantifies the intrinsic inadequacy of the approximate functional when evaluated on the exact density:
\begin{align}
  \Delta E^F_{\rm xc} &= E^{\rm DFA}[n^{\rm ref}] -  E[n^{\rm ref}] \nonumber \\
  &= E^{\rm DFA}_{\rm xc}[n^{\rm ref}] -  E_{\rm xc}[n^{\rm ref}],  
\end{align}  
where the second equality follows because all other terms in Eq.~\eqref{eq:energy-functional} are evaluated on the same density. The density-driven error captures the energetic consequence of using the approximate self-consistent density instead of the exact one:
\begin{equation}
    \Delta E^D = E^{\rm DFA}[n^{\rm DFA}]- E^{\rm DFA}[n^{\rm ref}].
\end{equation}  
This density error propagates through all terms of the energy functional:
\begin{equation}
    \Delta E^D = \Delta T_s + \Delta E_{\rm Hartree} + \Delta E^D_{\rm xc} + \int d{\bf r} \left( n^{\rm DFA}({\bf r}) - n^{\rm ref}({\bf r}) \right) v({\bf r}),  
\end{equation}
where the density-driven error in the xc energy is
\begin{equation}
\label{eq:density-driven-error-xc}
    \Delta E^D_{\rm xc} = E^{\rm DFA}_{\rm xc}[n^{\rm DFA}] - E^{\rm DFA}_{\rm xc}[n^{\rm ref}].
\end{equation}  

This framework provides a rigorous diagnostic tool: $\Delta E^F_{\rm xc}$ reveals the true quality of a functional's form, while $\Delta E^{D}_{\rm xc}$ quantifies how sensitive the xc energy is to density errors. In typical DFT calculations on atoms and molecules, $\Delta E^F_{\rm xc}$ dominates. This effect is what \citet{kim2013} termed ``normal'' behavior. However, in certain cases (which they called ``abnormal"), $\Delta E^{D}_{\rm xc}$ dominates, often signaled by unusually small HOMO-LUMO gaps. In such cases, the observed errors primarily reflect poor densities rather than poor functionals, and can be dramatically reduced by using more accurate reference densities. This approach led to the development of density-corrected DFT (DC-DFT)~\cite{Scuseria1992,Oliphant1994,Santra2021,song_density_2021}.

For solids, we leverage highly accurate quantum Monte Carlo (QMC) densities to define QMC-DFT, where $n^{\rm ref}({\bf r}) = n^{\rm QMC}({\bf r})$ and $E[n^{\rm ref}]=E^{\rm QMC}$. Recent advances in auxiliary-field QMC have made such reference densities available for crystalline systems~\cite{Chen2021}, enabling for the first time a systematic, material-specific analysis of functional versus density-driven errors in solids. In this work, we focus on the functional error $\Delta E^F_{\rm xc}$, which reflects the intrinsic quality of a functional independent of self-consistency artifacts, and the density-driven error in the xc energy, $\Delta E^D_{\rm xc}$, which quantifies how density inaccuracies propagate into energetic errors. Throughout this work, references to ``density-driven error" refer specifically to $\Delta E^D_{\rm xc}$ unless otherwise stated.

In addition to the ground-state density and the xc energy, an essential quantity for solids is the fundamental energy gap, or the quasiparticle band gap, $E_G$, defined as the difference between the ionization potential and the electron affinity. Within KS-DFT this quantity is often approximated by the KS band gap, $E^{\rm KS}_G = \varepsilon^{\rm LU} - \varepsilon^{\rm HO},$ i.e., the difference between the lowest unoccupied and highest occupied KS eigenvalues. While $E_G$ has a direct physical interpretation, $E^{\rm KS}_G$ is a property of an auxiliary noninteracting system and is not, in general, an observable of the interacting system. The two gaps are formally related: even with the exact xc functional, the fundamental gap differs from the KS gap due to the derivative discontinuity of the xc energy\cite{perdew_delta_xc1982, perdew_levy1983, sham_shluter1983, perdew_band_gap_problem1985, yang2000}, such that 
\begin{equation}
     E_G = E^{\rm KS}_G + \Delta_{\rm xc}.
\end{equation}
This derivative discontinuity term is missing in the semi-local approximations including LDA, GGA, and meta-GGA. In extended systems, it has been shown that the fundamental gap tends to the KS band gap, $E_G \to E^{\rm KS}_G$, within such approximations\cite{perdew_band_gap_problem1985, perdew2017_gks_gap}. Nevertheless, multiplicative KS potentials derived from LDA or GGA systematically underestimate experimental or quasiparticle gaps, often by up to a factor of two. In contrast, meta-GGAs introduce a dependence on the kinetic-energy density $\tau({\bf r})$, which renders the functional orbital dependent. Their xc potential is therefore non-multiplicative, and a consistent treatment requires the generalized Kohn–Sham (GKS) framework~\cite{seidl1996, gorling1997}. The additional orbital dependence effectively incorporates a degree of nonlocality, which increases the predicted band gap and recovers part of the missing discontinuity. As a result, meta-GGAs typically yield more realistic gaps, closer to the fundamental gap\cite{yang2016, labeda2024}. However, these gaps generally exceed the exact KS gap, so improved agreement with the fundamental gap does not imply a more accurate description at the KS level.

In this study, we analyze the KS band gap as a KS-system observable, on equal footing with the density and the xc energy, and decompose its error, analogously to the xc energy, into functional and density-driven contributions for multiplicative KS potentials, namely LDA and GGA. If we treat QMC results as near-exact benchmarks, the best xc functional would simultaneously reproduce the QMC density, xc energy, and KS band gap. We emphasize that, unlike \citet{borlido2019}, we do not assess functional performance with respect to the fundamental gap; rather, we use the KS band gap as a performance criterion for the quality of the functional at the KS level. 

For meta-GGAs, a meaningful comparison with the exact KS gap requires a multiplicative KS xc potential. We obtain this via KS inversion of the density corresponding to the orbital-dependent functional, yielding a potential expected to be numerically close to the optimized effective potential (OEP)\cite{talman1976, wu_yang2003, erhard2022}. This permits a direct comparison between the resulting KS band gap and the reference gap obtained from the QMC density. However, because the inverted potential is not a density functional, a decomposition of the total error into functional and density-driven components is not possible in this case.

We apply this analysis to three prototypical solids representing distinct bonding environments: silicon (covalent semiconductor), sodium chloride (ionic insulator), and copper (metal). Our results reveal material-dependent patterns of error cancellation, identify functionals with minimal intrinsic errors for each material class, and uncover cases where density-driven errors dominate, suggesting opportunities for substantial improvement through density correction. These findings provide practical guidance for functional selection and carry important implications for emerging applications such as DFT-based machine learning potentials, where systematic density errors may compromise transferability across materials.

\section{Methods and Computational Details}

An ideal functional should minimize both the functional error, $\Delta E^F_{\rm xc}$, and the density-driven error, $\Delta E^D_{\rm xc}$. In typical DFT calculations on atoms and molecules, $\Delta E^F_{\rm xc}$ dominates over $\Delta E^D_{\rm xc}$ in magnitude~\cite{kim2013}, as densities are relatively insensitive quantities that remain close to their reference counterparts even when approximate functionals are used. However, a small $\Delta E^D_{\rm xc}$, as defined in Eq.~\eqref{eq:density-driven-error-xc}, can arise for two distinct reasons. The functional may be insufficiently sensitive to density variations, unable to distinguish between approximate and reference densities, in this case, the functional form itself limits accuracy. Alternatively, the approximate functional may successfully guide the self-consistent iterations toward a density very close to the reference one, leading to small density-driven error regardless of its approximate form quality. These scenarios can be distinguished by further examining the root-mean-square difference (RMSD) between the approximate and QMC densities:
\begin{equation} 
{\rm RMSD}_{[n({\bf r})]} = \sqrt{ \frac{1}{N} \sum_{i=1}^{N} \left( n^{\rm DFA}({\bf r}_i) - n^{\rm QMC}({\bf r}_i) \right)^2 },
\label{eq:RMSD_n}
\end{equation}  
where $N$ is the number of real-space grid points, and ${\bf r}_i$ denotes the position of the $i$-th grid point. A small RMSD indicates good density agreement, while a large RMSD signals substantial density errors that contribute to $\Delta E^D_{\rm xc}$.

Besides the xc energy decomposition, we also analyze the total xc energy error:
\begin{equation}
  \Delta E_{\rm xc} = E^{\rm DFA}_{\rm xc}[n^{\rm DFA}] - E^{\rm QMC}_{\rm xc}, 
\end{equation}
where $E^{\rm QMC}_{\rm xc}$ is the QMC reference xc energy~\cite{Aouina2024,Chen2021}. This total error reflects the sum of both contributions: $\Delta E_{\rm xc} = \Delta E^F_{\rm xc} + \Delta E^D_{\rm xc}$. Comparing $\Delta E_{\rm xc}$ with $\Delta E^F_{\rm xc}$ reveals the extent of error cancellation between functional and density-driven components.

For KS band gaps, we define functional errors analogously. For the indirect KS band gap and the direct KS band gap at $\Gamma$, these are, respectively
\begin{align}
    \Delta E_{\rm IndG}^{F} &= E_{\rm IndG}^{\rm DFA}[n^{\rm QMC}] - E_{\rm IndG}^{\rm QMC},  \\
    \Delta E_{\rm DG}^{F} &= E^{\rm DFA}_{\rm DG}[n^{\rm QMC}] - E_{\rm DG}^{\rm QMC}, 
\end{align}
where $E_{\rm IndG}^{\rm QMC}$ and $E_{\rm DG}^{\rm QMC}$ are the near-exact KS band gaps obtained from QMC densities~\cite{Aouina2023}. The corresponding density-driven errors, $\Delta E^{D}_{\rm IndG}$ and $\Delta E^{D}_{\rm DG}$, are defined analogously to Eq.~\eqref{eq:density-driven-error-xc}, as 
\begin{align}
    \Delta E_{\rm IndG}^{D} &= E_{\rm IndG}^{\rm DFA}[n^{\rm DFA}] -E_{\rm IndG}^{\rm DFA}[n^{\rm QMC}],  \\
    \Delta E_{\rm DG}^{D} &= E^{\rm DFA}_{\rm DG}[n^{\rm DFA}] -E^{\rm DFA}_{\rm DG}[n^{\rm QMC}].
\end{align}

All energy errors are reported both in absolute values (eV) and on a relative scale normalized to the LDA-PW92 error, allowing direct comparison of functional performance across different materials and properties. 

We performed both self-consistent DFT and one-shot QMC-DFT calculations using the \texttt{Density-Functional ToolKit (DFTK.jl)}~\cite{DFTKjcon} with \texttt{Julia} bindings to Libxc~\cite{marques2012,lehtola2018, tran2026}. We evaluated 70 functionals spanning three rungs of Jacob's ladder: local density approximations (LDA), generalized gradient approximations (GGA), and meta-generalized gradient approximations (meta-GGA).  

To ensure direct compatibility and fair comparison with QMC benchmarks, we adopted identical calculation parameters to those used in Ref.~\onlinecite{Chen2021}: the same norm-conserving pseudopotentials, plane-wave kinetic energy cutoffs, and k-point grids for each material. This consistency eliminates systematic differences between our DFT results and the QMC reference data. Full details of the computational parameters, along with the complete list of functionals and their references, are provided in the Supplementary Material. 

For meta-GGA functionals, the xc energy depends not only on the density and its gradient but also on the kinetic energy density $\tau({\bf r})$, which is constructed from a set of GKS orbitals $\{\phi_i^{\rm GKS}({\bf r})\}$ that minimize the given DFA. A rigorous estimation of the functional error requires inverting the GKS equations to obtain the orbitals $\{\phi_i^{\rm GKS}[n^{\rm QMC}]\}$ that reproduce the QMC density. The functional error can then be expressed as
\begin{equation}
\Delta E^F_{\rm xc} = E^{\rm DFA}_{\rm xc}[\{\phi_i^{\rm GKS}[n^{\rm QMC}]\}] - E^{\rm QMC}_{\rm xc}.
\end{equation}
However, GKS inversion is significantly more challenging than standard KS inversion and remains largely unexplored in the literature~\cite{Khanna2026}. In this work,  we choose to approximate the functional error using the scheme discussed in Ref.~\citenum{ravindran2025_preprint}. The error is first decomposed as
\begin{equation}
\Delta E^F_{\rm xc} = \Delta E^{\rm NL}_{\rm DFA}[n^{\rm QMC}] + \left(E^{\rm DFA}_{\rm xc}[\{\phi_i^{\rm KS}[n^{\rm QMC}]\}] - E^{\rm QMC}_{\rm xc}\right),
\end{equation}
where
\begin{equation}
\Delta E^{\rm NL}_{\rm DFA}[n^{\rm QMC}] = E^{\rm DFA}_{\rm xc}[\{\phi_i^{\rm GKS}[n^{\rm QMC}]\}] - E^{\rm DFA}_{\rm xc}[\{ \phi_i^{\rm KS}[n^{\rm QMC}]\} ] ,
\end{equation}
quantifies the difference between the GKS and KS representations of the same density. It can also be interpreted as the DFA's ``nonlocality" energy at the QMC density~\cite{ravindran2025}, and it vanishes for LDAs and GGAs. We then approximate this term as
\begin{equation}
\Delta E^{\rm NL}_{\rm DFA}[n^{\rm QMC}] \approx E^{\rm DFA}_{\rm xc}[\{\phi_i^{\rm GKS}[n^{\rm DFA}]\}] - E^{\rm DFA}_{\rm xc}[\{\phi_i^{\rm KS}[n^{\rm DFA}]\}] \leq 0,
\end{equation}
where the KS orbitals $\{\phi_i^{\rm KS}[n^{\rm DFA}]\}$ are obtained through KS inversion of the self-consistent DFA density $n^{\rm DFA}$. 

The situation is more challenging for band gaps. Not only do we not have the GKS orbitals for the QMC density, as this would require the GKS inversion, but the GKS band gaps are also not directly comparable to the KS QMC reference. This is because GKS functionals target the experimental band gap, and it has been shown that their band gaps capture a portion of the xc derivative discontinuity $\Delta_{\rm xc}$~\cite{seidl1996, perdew2017_gks_gap}, which is absent from the exact KS gap~\cite{perdew2017_gks_gap}. To address this, for each meta-GGA, we invert the self-consistent density to obtain a local exchange-correlation (LXC) potential, $v^{\rm LXC}_{\rm xc}({\bf r})$, and extract the band gap from the resulting KS eigenvalues. A similar study of LXC potentials of non-local functionals can be found in Ref.~\citenum{ravindran2025}. Since the potential $v^{\rm LXC}_{\rm xc}({\bf r})$ is a function rather than a functional, we are not able to separate the functional and density-driven errors as done above for LDAs and GGAs; we therefore restrict ourselves to estimating the total error $\Delta E_{\rm DG} = \Delta E^{F}_{\rm DG} + \Delta E^{D}_{\rm DG}$, and similarly for the indirect gap $\Delta E_{\rm IndG}$.

Finally, since copper is metallic, the absence of a band gap calls for an alternative descriptor of the electronic structure. We use the d-band width, $W_d$, defined, following Ref.~\citenum{chen_dband_width2025}, as the root-mean-square deviation of the d-band energy weighted 
by the projected d-density of states $g_d(E)$
\begin{equation}
    W_d^2 = \frac{\int_{-\infty}^{+\infty} g_d(E)(E - E_d)^2 \, dE}{\int_{-\infty}^{+\infty} g_d(E) \, dE},
    \label{eq:dband_width}
\end{equation}
where $E_d$ is the average energy of the projected d-density of states,
\begin{equation}
    E_d = \frac{\int_{-\infty}^{+\infty} g_d(E)(E - E_F) \, dE}{\int_{-\infty}^{+\infty} g_d(E) \, dE},
    \label{eq:Ed}
\end{equation}
and $E_F$ is the Fermi level. As explained for the band gap, we are not able to separate the functional and density-driven errors here either for meta-GGA functionals, and we therefore can compute only the total error on the d-band width for these functionals.

\section{Assessing Functionals Using Functional Error}

As demonstrated by \citet{kim2013}, DFAs should be re-evaluated by distinguishing between the intrinsic error of the functional form (measured by evaluating the functional on exact densities) and density-driven errors, which depend on both the system and the property under investigation. In this section, we systematically assess the intrinsic quality of xc functionals by computing their functional errors for three representative solids. Using QMC densities as reference, we perform one-shot KS calculations to evaluate each approximate functional at the exact density, yielding energies and KS eigenvalues free from self-consistency artifacts. This approach reveals the true quality of each functional independent of its ability to produce accurate densities.

A conceptually similar study by \citet{mezei_electron_2017} analyzed functional and density-driven errors for a large set of molecules and atoms. However, their density-driven errors were computed using the PBE functional as reference, rather than following the original definition of \citet{kim2013} which requires exact densities. This approach conflates errors from two approximate functionals, making it difficult to assess the intrinsic quality of each functional independently. In contrast, the availability of high-quality QMC benchmarks for solids~\cite{Chen2021} enables us to follow the original framework rigorously, using truly accurate reference densities. This provides more reliable conclusions about functional and density-driven errors in crystalline materials, where bonding environments differ fundamentally from molecular systems.

\subsection{Silicon}

For silicon, the top-performing functionals for the xc energy exhibit remarkably small functional errors (Table~\ref{tab:Ef_si}). GP86, SOGGA11, BP86, and PW91P86 achieve errors ranging from only 7\% to 12\% of the LDA value. Notably, three of these four functionals incorporate P86 correlation, suggesting that this component plays a crucial role in reducing xc energy errors for the covalent semiconductor. The P86 correlation functional is built to capture the correlation energy of the homogeneous electron gas (HEG) and inhomogeneity effects beyond the random phase approximation recovering the density-gradient expansion of the correlation energy in the slowly varying limit\cite{perdew_correlation1986}. This excellent performance is particularly impressive given that these top-performing functionals were not empirically optimized for a class of materials. 

Most functionals studied here follow the typical pattern identified by \citet{kim2013} and \citet{nam_measuring_2020} for molecules: the functional error dominates over the density-driven error in magnitude, $|\Delta E^F_{\rm xc}| > |\Delta E^D_{\rm xc}|$. This ``normal'' behavior indicates that improving the functional form would yield greater benefits than improving the density. However, several functionals exhibit the opposite, ``abnormal'' behavior where $|\Delta E^D_{\rm xc}| > |\Delta E^F_{\rm xc}|$. The empirically fitted SOGGA11 and $\tau$-HCTH show density-driven errors approximately two times larger than their functional errors. For these functionals, evaluating on the QMC density reduces the total error by a factor of three, demonstrating substantial improvement through density correction alone. Similarly, HCTH407, predecessor of $\tau$-HCTH, shows a functional error contributing less than half of its total xc energy error, again indicating that poor self-consistent density is the primary source of inaccuracy. 

An interesting case of error cancellation occurs with BP86, where the density-driven error slightly exceeds the functional error in magnitude but carries the opposite sign. This fortuitous cancellation yields a total xc energy within 0.01 eV of the QMC reference. While this produces excellent results for silicon, such error cancellation is system-dependent and may not transfer to other materials or properties.

Turning to KS band gaps (Table~\ref{tab:EDG_f_si}), we focus on functionals that outperform LDA for both the direct gap at $\Gamma$ and the indirect gap, evaluating their performance based on functional errors. 

The GGA functional EDF1 ranks first, providing excellent accuracy for the direct band gap but showing larger errors for the indirect gap. ZLP-LDA offers more balanced performance across both gaps, though its xc energy accuracy is inferior to LDA(PW92).

Encouragingly, the functionals that excel for xc energies also perform well for band gaps. BP86, PW91P86, and GP86 all maintain functional errors below 0.05 eV for both direct and indirect gaps when evaluated on the QMC density. This consistency across properties suggests genuine functional quality rather than fortuitous error cancellation for a single observable.

Several functionals show abnormal behavior for band gaps, with density-driven errors exceeding functional errors:
EDF1 for the direct gap, and BPW91 and revPBE for the indirect gap. For these functionals, density correction schemes will be beneficial in improving the description of the KS system.

Among the top four functionals for xc energy, all except SOGGA11 demonstrate consistent excellence across both band gaps and xc energy when evaluated on QMC densities. GP86 emerges as the most balanced performer, ranking 1st for xc energy, 2nd for the indirect gap, and 5th for the direct gap. This combination of strong performance across multiple properties and small functional errors makes GP86 particularly reliable for silicon.

\begin{figure}
    \centering
    \includegraphics[width=0.8\linewidth]{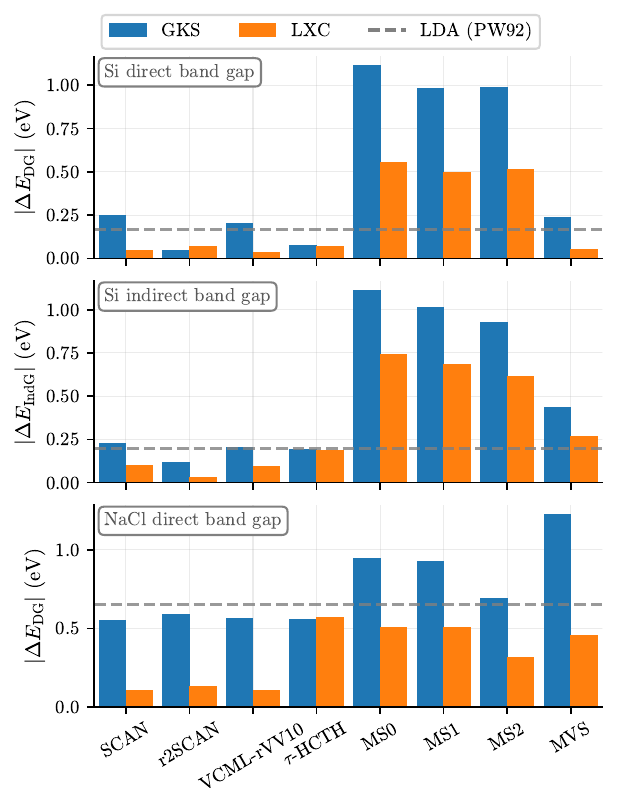}
    \caption{Absolute deviations of the direct and indirect band gaps of Si and the direct band gap of NaCl from the KS QMC reference, computed with eight meta-GGA functionals. Results are shown for both the GKS band gaps and the corresponding gaps from the local exchange-correlation (LXC) potential, obtained by inverting the self-consistent GKS density. The larger GKS gaps deviations are expected, since GKS gaps are not strictly comparable to KS QMC gaps. In fact, the GKS functionals are designed to reproduce experimental gaps, which capture part of the derivative discontinuity $\Delta_{\rm xc}$. However, their corresponding LXC gaps show systematically smaller deviations from the KS QMC reference. The dashed line marks the deviation of the LDA-PW92 functional, serving as a baseline for comparison. } 
    \label{fig:errors_band_gap_GKS_LXC}
\end{figure}

Finally for the meta-GGAs, as shown in Fig.~\ref{fig:errors_band_gap_GKS_LXC}, the inverted LXC potential generally makes the band gaps closer to the exact KS band gaps. These results are expected as discussed in the literature\cite{yang2016, perdew2017_gks_gap, ravindran2025}, while here, thanks to the QMC benchmark, we can give numerical evidence for it. An exception arises for r2SCAN in the case of the direct band gap, which shows a slightly larger gap in its LXC version than the GKS gap, while it maintains the expected behaviour in the case of the indirect gap. For the functional $\tau$-HCTH, for which the density-driven error in the xc energy dominates, the LXC potential brings the gaps only slightly toward the exact gaps, but still very similar to the GKS gaps. Among all these meta-GGA functionals, VCML-rVV10 appears the best in providing consistently high-quality density, small functional and total xc energy errors, as shown in Table~\ref{tab:Ef_si}, and LXC gaps closer to the QMC KS gaps.

\begin{table}
\centering
\caption{Top ten functionals yielding the smallest functional error for Si. The different columns contain the functional ranking, the rung, the publication year, the functional error $\Delta E^F_{\rm xc}$ in eV, the density-driven error $\Delta E^D_{\rm xc}$ in eV, the LDA-normalized functional error $\Delta E^F_{\rm xc, norm}$ in \%, the ranking with respect to the error in the total $\Delta E_{\rm xc}$ and the ranking with respect to the RMSD  between  the approximate and QMC densities.}
\label{tab:Ef_si}
\begin{tabular}{llrrrrrr}
\toprule
 \# & functional & rung  & $\Delta E^{F}_{\rm xc}$ & $\Delta E^D_{\rm xc}$ & $\Delta E^F_{\rm xc, norm}$ & \#$\Delta E_{\rm xc}$ & \#RMSD$_{[n({\bf r})]}$ \\
\midrule
1 & GP86 & GGA & -0.081 & -0.068 & 7.23 & 3 & 5 \\
2 & SOGGA11 & GGA  & 0.096 & 0.201 & 8.54 & 7 & 51 \\
3 & BP86 & GGA  & 0.101 & -0.112 & 8.96 & 1 & 8 \\
4 & PW91P86 & GGA & 0.137 & -0.092 & 12.13 & 2 & 6 \\
5 & $\tau$-HCTH & mGGA  & -0.223 & -0.446 & 19.82 & 22 & 47 \\
6 & GPW91 & GGA & 0.265 & -0.016 & 23.48 & 6 & 7 \\
7 &  VCML-rVV10 & mGGA  & 0.319 & -0.105 & 28.32 & 4 & 2 \\ 
8 & GPBE & GGA & 0.331 & 0.014 & 29.34 & 8 & 9 \\
9 & PBEP86 & GGA & 0.333 & -0.086 & 29.59 & 5 & 16 \\
10 & BPW91 & GGA  & 0.447 & -0.060 & 39.67 & 9 & 13 \\
\bottomrule
\end{tabular}
\end{table}

\begin{table}
\setlength{\tabcolsep}{3pt}
\centering
\caption{Top ten functionals yielding the smallest functional error for the indirect and direct KS band gap of Si, from the subset that outperforms LDA-PW92 for both gaps. The different columns contain the functional ranking with respect to the direct KS band gap, the rung, the publication year, the functional error $\Delta E^F_{\rm DG}$ and the density-driven error $\Delta E^D_{\rm DG}$ in eV for the direct KS band gap, the functional error $\Delta E^F_{\rm IndG}$ and the density-driven error $\Delta E^D_{\rm IndG}$ in eV for the indirect KS band gap, 
the ranking with respect to the functional error (see Table~\ref{tab:Ef_si}) and the ranking with respect to the RMSD  between  the approximate and QMC densities.}
\label{tab:EDG_f_si}
\begin{tabular}{llrrrrrrrr}
\toprule
\# & functional & rung  & $\Delta E^{F}_{\rm DG}$ & $\Delta E^{D}_{\rm DG}$ & $\Delta E^{F}_{\rm IndG}$ & $\Delta E^{D}_{\rm IndG}$  & \#$\Delta E^F_{\rm xc}$ & \#RMSD$_{[n({\bf r})]}$ \\
\midrule

1  & EDF1 & GGA &  -0.021 & -0.029 & 0.174 & -0.012  & 38 & 22 \\
2  & ZLP-LDA & LDA & -0.038 & -0.007 & 0.026 & -0.011  & 41 & 4 \\
3  & BP86 & GGA &  -0.041 & -0.011 & 0.050 & -0.011  & 3 & 8 \\
4  & PW91P86 & GGA &  -0.042 & -0.009 & 0.038 & -0.010  & 4 & 6 \\
5  & GP86 & GGA &  -0.044 & -0.004 & 0.012 & -0.002  & 1 & 5 \\
6  & PBEP86 & GGA &  -0.072 & -0.008 & 0.029 & -0.005 & 8 & 16 \\
7  & BPW91 & GGA &  -0.077 & -0.003 & -0.002 & 0.003 & 10 & 13 \\
8  & OPTX & GGA &  -0.079 & -0.032 & 0.176 & 0.049 &  61 & 36 \\
9  & PW91 & GGA &  -0.079 & -0.001 & -0.014 & 0.004 & 12 & 10 \\
10 & GPW91 & GGA &  -0.081 & 0.005 & -0.040 & 0.011 & 6 & 7 \\
\bottomrule
\end{tabular}
\end{table}

\subsection{Sodium Chloride}

For NaCl, all ten functionals with the smallest functional errors are GGAs (Table~\ref{tab:Ef_nacl}). PBEPW91 achieves exceptional accuracy, with a functional error of only 1.49\% relative to LDA. The next best performers show progressively larger errors: PBEP86 at 4.45\%, followed by PW91PBE and PBE both below 6\%. The prevalence of PBE-based functionals among top performers is noteworthy and can be understood by considering that NaCl, as an ionic insulator, has more localized electron density than silicon. This makes the system more similar to molecules and atoms, where PBE is known to perform well~\cite{Shahi2019}.

Most functionals follow the normal pattern where functional errors dominate over density-driven errors. Only SOGGA11 shows abnormal behavior, with its density-driven error exceeding the functional error. Given NaCl's larger KS band gap compared to Si, this finding aligns with observations by \citet{kim2013} for finite systems, where large KS gaps typically correlate with normal behavior: manifesting in density-driven errors less pronounced. When evaluated on the QMC density, SOGGA11's xc energy error is reduced by a factor of two, demonstrating the benefit of density correction.

Interestingly, for most top performers, the relationship between total error $\Delta E_{\rm xc}$ and functional error $\Delta E^F_{\rm xc}$ is not straightforward. Using the QMC density improves xc energy predictions for PBEPW91 and PW91PBE, but actually worsens them for PBEP86 and PBE due to unfavorable error cancellation in their self-consistent calculations. This highlights that good performance with self-consistent densities may sometimes result from fortuitous cancellation rather than intrinsic functional quality.

For KS band gaps, the top three functionals for xc energy show only moderate performance, underestimating the band gap by 0.19 to 0.37 eV. The GGA functional mPWPBE offers more balanced performance, with normalized-LDA errors of 11.06\% for xc energy and 9.24\% for the band gap. Focusing exclusively on band gap accuracy reveals a different set of top performers: PBEOP, GP86, BPW91, and BPBE achieve errors below 5 meV. However, this excellent band gap accuracy stems primarily from density correction effects rather than intrinsic functional quality. For these functionals, the density-driven error in the direct band gap, $\Delta E_{\rm DG}^D$, is 4 to 23 times larger than the functional error $\Delta E_{\rm DG}^F$, with GP86 showing the most extreme ratio. This indicates that these functionals are abnormally sensitive to density quality for band structure predictions, and density correction could largely improve their performance.

Among the functionals with excellent band gap accuracy, BPBE shows the best overall balance, with a functional error of approximately 0.75 eV (12.59\% of the LDA error) for the xc energy while maintaining less than 5 meV band gap errors. This combination makes it a promising choice for NaCl when accurate densities are available, though its self-consistent results are less impressive due to the absence of error cancellation.

The comparison of GKS and LXC gaps in Fig.~\ref{fig:errors_band_gap_GKS_LXC} shows a similar trend as for Si, with LXC gaps overall smaller and closer to the QMC KS gaps. VCML-rVV10, together with SCAN and r2SCAN, yields very accurate KS gaps and shows the expected decrease when using the LXC potential rather than the GKS framework. Again, $\tau$-HCTH behaves unexpectedly, yielding this time a GKS gap slightly larger than the LXC gap. Finally, the exchange-only functionals MS0, MS1, MS2, and MVS, when considered in their corresponding LXC framework, yield band gaps better than the LDA-PW92 gap. This improvement, which was absent in the case of Si, is a consequence of the better density quality of these functionals for NaCl than for Si.

\begin{table}
\centering
\caption{Top ten functionals yielding the smallest functional error for NaCl. The errors $\Delta E^F_{\rm xc}$ and $\Delta E^D_{\rm xc}$ are in eV, and $\Delta E^F_{\rm xc, norm}$ is in \%. Column definitions are the same as in Table~\ref{tab:Ef_si}.}
\label{tab:Ef_nacl}
\begin{tabular}{llrrrrrr}
\toprule
 \# & functional & rung  & $\Delta E^{F}_{\rm xc}$ & $\Delta E^D_{\rm xc}$ & $\Delta E^F_{\rm xc, norm}$ & \#$\Delta E_{\rm xc}$ & \#RMSD$_{[n({\bf r})]}$ \\
\midrule
1 & PBEPW91 & GGA &  -0.089 & -0.061 & 1.49 & 2 & 5 \\
2 & PBEP86 & GGA & -0.264 & 0.136 & 4.45 & 1 & 11 \\
3 & PW91PBE & GGA &  -0.306 & -0.095 & 5.15 & 4 & 9 \\
4 & PBE & GGA &  0.317 & -0.045 & 5.33 & 3 & 7 \\
5 & mPWPBE & GGA &  -0.657 & -0.264 & 11.06 & 9 & 16 \\
6 & PW91 & GGA &  -0.711 & -0.111 & 11.97 & 8 & 8 \\
7 & PBELYP1W & GGA &  0.735 & -0.022 & 12.39 & 5 & 13 \\
8 & BPBE & GGA &  -0.747 & -0.389 & 12.59 & 10 & 22 \\
9 & PW91P86 & GGA &  -0.887 & 0.086 & 14.93 & 7 & 15 \\
10 & revPBE & GGA &  -1.011 & -0.554 & 17.03 & 16 & 23 \\
\bottomrule
\end{tabular}
\end{table}

\begin{table}
\centering
\caption{Top ten functionals yielding the smallest functional error for the direct KS band gap of NaCl. The errors $\Delta E^F_{\rm DG}$ and  $\Delta E^D_{\rm DG}$ are in eV. Refer to Table~\ref{tab:EDG_f_si} for column definitions.}
\label{tab:EDG_f_nacl}
\begin{tabular}{llrrrrr}
\toprule
 \# & functional & rung  & $\Delta E^{F}_{\rm DG}$ & $\Delta E^{D}_{\rm DG}$ & \#$\Delta E^{F}_{\rm xc}$ & \#RMSD$_{[n({\bf r})]}$ \\
\midrule
1 & PBEOP & GGA & 0.001 & -0.006 & 30 & 2 \\
2 & GP86 & GGA  & 0.002 & 0.057 & 21 & 27 \\
3 & BPW91 & GGA  & -0.003 & 0.022 & 11 & 21 \\
4 & BPBE & GGA  & -0.004 & 0.028 & 8 & 22 \\
5 & GLYP & GGA  & 0.024 & 0.008 & 27 & 14 \\
6 & BECKE & GGA  & 0.062 & 0.028 & 63 & 1 \\
7 & mPWPBE & GGA  & -0.085 & 0.036 & 5 & 16 \\
8 & N12 & GGA  & -0.122 & -0.057 & 20 & 54 \\
9 & revPBE & GGA  & 0.132 & -0.001 & 10 & 23 \\
10 & PW91X & GGA  & -0.134 & 0.064 & 64 & 3 \\
\bottomrule
\end{tabular}
\end{table}

\subsection{Copper}
Metallic copper presents a significantly greater challenge for density functionals than either silicon or sodium chloride. Only five functionals achieve functional errors smaller than LDA (Table~\ref{tab:Ef_cu}), and even the best performer, SOGGA, reaches only 67.74\% of the LDA error. This stark contrast with the covalent and ionic systems highlights fundamental limitations in current semilocal approximations for metallic bonding.

PBEsol ranks second with a functional error of 71.10\% relative to LDA. Interestingly, PBEsol exhibits large error cancellation between functional and density-driven errors with opposite signs, making it superior to SOGGA when used with self-consistent densities despite having a larger intrinsic functional error. The remaining three functionals that outperform LDA are PBELYP, VWN (an LDA variant), and PBEOP. The success of PBELYP and PBEOP suggests that combining PBE exchange with either LYP or OP correlation is particularly effective for the xc energy of metallic systems, the same trend has been observed for the density\cite{Aouina2024}, though even these combinations struggle to achieve accuracies comparable to what GGAs deliver for semiconductors and insulators. 

This result can be compared with the conclusions of Paier \textit{et al.}~\cite{Paier2007},
who attributed roughly two-thirds of the B3LYP error in atomization energies of metals to
the LYP correlation functional rather than to the exchange part.
In that work, the authors acknowledged that disentangling whether the error originates from
an inadequate description of the metallic limit or of the atomic limit is difficult.
Our results suggest that the atomic limit may be the more critical factor: the comparatively
good performance of PBELYP and PBEOP for the xc energy of copper, despite LYP's known
deficiency in the HEG limit, indicates that the error cancellation observed in B3LYP for
metals is more likely rooted in a poor description of the atomic reference state than in the
treatment of the metallic bulk.

The dominance of GGAs and LDAs among the top performers contrasts sharply with silicon and sodium chloride, where more sophisticated functionals showed promise. All top-10 performers for copper are either GGAs or LDA variants (VWN, PW92, PZ81), with no meta-GGAs achieving competitive accuracy. More concerning is that even the best functionals show normalized errors ranging from 67\% to 164\% of the LDA value. This indicates that climbing Jacob's ladder offers minimal benefit for metals, a limitation that has persisted despite decades of functional development.

The difficulty in describing metallic copper likely stems from the delocalized nature of the valence electrons and the importance of electron correlation at the Fermi surface. Standard semilocal approximations, which rely on local or semi-local density information, may be fundamentally inadequate for capturing the long-range and collective electronic behavior characteristic of metals. This suggests that further improvements for metallic systems may require going beyond semilocal approximations to include nonlocal correlation or exact exchange contributions. 

Nearly all functionals examined for copper exhibit normal behavior, with functional errors dominating over density-driven errors. The sole exception is the LDA variant VWN, which shows a density-driven error, $\Delta E^D_{\rm xc}$, exceeding its functional error. This pattern indicates that for metallic systems, the primary limitation lies in the functional approximation itself rather than the quality of self-consistent densities. Unlike silicon, where several functionals showed abnormal behavior suggesting potential benefits from density correction, copper's errors are predominantly functional-driven. This implies that improving predictions for metals requires developing better functionals rather than improving density quality.

The absence of a detailed band gap analysis for copper reflects its metallic character, with no band gap to evaluate. However, an alternative descriptor for the metal is the d-band width $W_d$, defined in Eq.~\eqref{eq:dband_width}. Table~\ref{tab:Wd_f_cu} reports the ten functionals with the smallest $W_d$ functional errors, and the picture
 differs qualitatively from the xc energy analysis. The functional errors are at most a few meV across the full set, while the density-driven errors are systematically two to three orders of magnitude larger. SOGGA11 is the extreme case, with a functional error magnitude of 0.04 meV against a density-driven error magnitude of 11 meV. The same hierarchy holds for most functionals we tested, not only those listed in the table. Once the density is fixed to the QMC reference, the d-band width $W_d$ becomes largely insensitive to which xc functional is used to evaluate it. In copper, $W_d$ is density-limited rather than functional-limited, the reverse of what we found for the xc energy. Density correction alone would therefore eliminate most of the $W_d$ error for most  of the functionals examined here.

\begin{table}
\centering
\caption{Top ten functionals yielding the smallest functional error for Cu. The errors $\Delta E^F_{\rm xc}$ and $\Delta E^D_{\rm xc}$ are in eV, and $\Delta E^F_{\rm xc, norm}$ is in \%. The horizontal line separates functionals that outperform LDA (above) from those that do not (below). Column definitions are the same as in Table~\ref{tab:Ef_si}.}
\label{tab:Ef_cu}
\begin{tabular}{llrrrrrr}
\toprule
 \# & functional & rung &  $\Delta E^{F}_{\rm xc}$ & $\Delta E^D_{\rm xc}$ &$\Delta E^F_{\rm xc, norm}$ & \#$\Delta E_{\rm xc}$ & \#RMSD$_{[n({\bf r})]}$ \\
\midrule
1 & SOGGA & GGA  & 3.966 & 1.707 & 67.74 & 3 & 30 \\
2 & PBEsol & GGA  & -4.163 & 1.480 & 71.10 & 2 & 28 \\
3 & PBELYP & GGA  & -4.580 & -2.259 & 78.22 & 4 & 7 \\
4 & VWN & LDA  & 5.091 & 5.452 & 86.95 & 6 & 49 \\
5 & PBEOP & GGA  & -5.164 & -2.878 & 88.20 & 5 & 14 \\
\midrule
6 & PW92 & LDA  & 5.855 & 5.499 & 100.00 & 8 & 50 \\
7 & MOHLYP & GGA  & 6.042 & -3.677 & 103.20 & 1 & 41 \\
8 & PZ81 & LDA  & 6.655 & 5.591 & 113.67 & 9 & 51 \\
9 & PW91LYP & GGA  & -8.454 & -2.229 & 144.40 & 7 & 3 \\
10 & BLYP & GGA  & -9.564 & -2.705 & 163.34 & 10 & 8 \\
\bottomrule
\end{tabular}
\end{table}

\begin{table}
\centering
\caption{Top ten functionals yielding the smallest functional error in the d-band width $W_d$ for Cu. The functional error $\Delta W_d^F$ and density-driven error $\Delta W_d^D$ are in eV. Column definitions are the same
   as in Table~\ref{tab:Ef_si}.} 
\label{tab:Wd_f_cu}
\begin{tabular}{llrrrrr}
\toprule
 \# & functional & rung  & $\Delta W_d^F (\times 10^{-4})$ & $\Delta W_d^D (\times 10^{-4})$ &  \#$\Delta E^{F}_{\rm xc}$ & \#RMSD$_{[n({\bf r})]}$ \\
\midrule
1 & SOGGA11 & GGA  & -0.441 & -110.261 & 33 & 27 \\
2 & PBEPW91 & GGA  & -0.973 & -72.818 & 21 & 24 \\
3 & PW91P86 & GGA  & -4.050 & -62.624 & 29 & 11 \\
4 & MOHLYP & GGA  & 4.157 & -71.550 & 7 & 41 \\
5 & PBEPZ81 & GGA  & -4.859 & -325.309 & 43 & 32 \\
6 & PW91PZ81 & GGA & 5.601 & -342.721 & 46 & 29 \\
7 & OPW91 & GGA  & -6.288 & -353.404 & 52 & 61 \\
8 & GP86 & GGA  & -6.593 & -66.788 & 35 & 13 \\
9 & OPBE & GGA & -6.709 & -306.571 & 51 & 60 \\
10 & BPBE & GGA  & 8.356 & -110.251 & 25 & 25 \\
\bottomrule
\end{tabular}
\end{table}

\begin{figure}
    \centering
    \includegraphics[width=\linewidth]{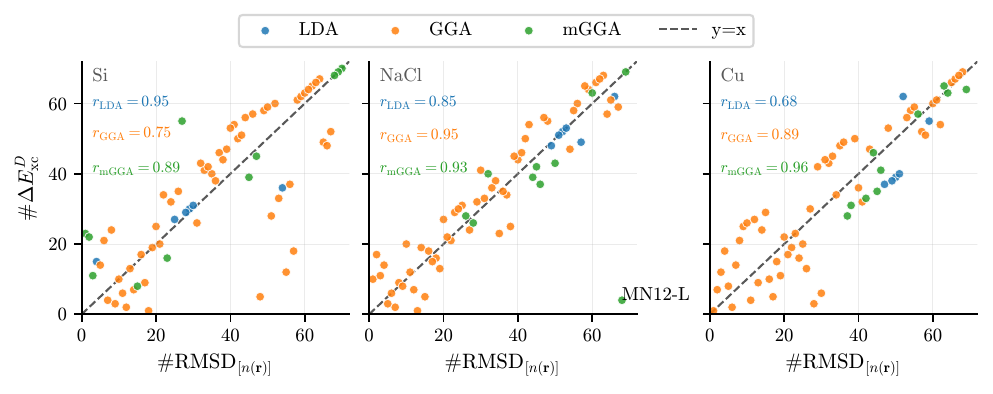}
    \caption{Ranking of functionals based on density-driven error, $\Delta E^D_{\rm xc}$, versus their ranking based on density error, RMSD$_{[n({\bf r})]}$. Each point represents a functional, illustrating the correlation between density quality and density-driven error across the tested systems. Pearson correlation coefficients $r$  computed separately for each functional rung are shown in the top-left corner of each panel. MN12-L is excluded from the NaCl panel.} 
    \label{fig:correlation_dde_rmsd}
\end{figure}

\begin{figure}
    \centering
    \includegraphics[width=\linewidth]{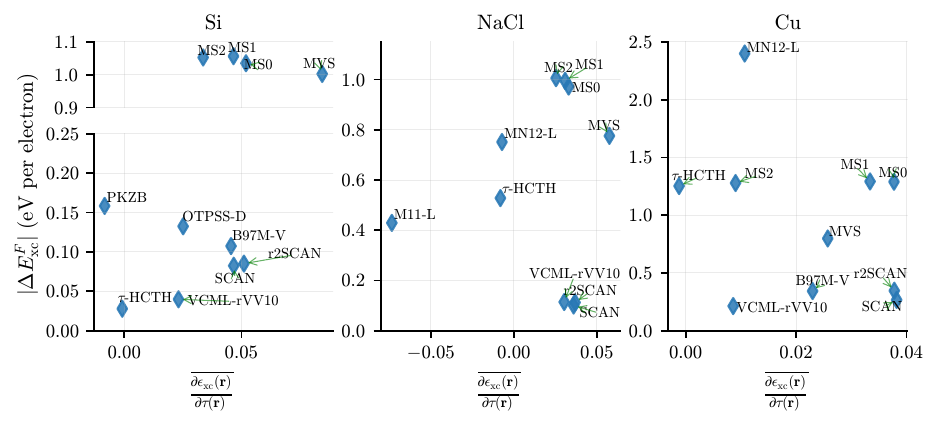}
  \caption{Comparison of absolute xc energy functional errors $|\Delta E_{\text{xc}}^F|$ (eV per electron) against the average of the derivative of the xc energy density with respect to the kinetic energy density, $\overline{\frac{\partial \epsilon_{\rm xc}({\bf r})}{\partial \tau({\bf r})}}$, for Si, NaCl, and Cu. Green arrows indicate the mapping of functional names to their respective data points.}
    \label{fig:ultra-nonlocality}
\end{figure}

\subsection{Correlation between density and density-driven errors} 
Figure~\ref{fig:correlation_dde_rmsd} compares the rankings of functionals based on their density-driven error and their overall density error. A positive correlation is observed across all three systems, indicating that, for the functionals considered, improved density accuracy typically translates into lower density-driven errors. This correlation was not observed in the study by~\citet{mezei_electron_2017}, which likely stems from a difference in the definition used: while they calculated the density-driven error in the PBE xc energy, our approach uses the proper, functional-independent definition~\cite{kim2013}, which is not influenced by the specific form of the functional used to evaluate the density.

Although there is an overall correlation, its strength and structure vary across materials. To quantify this, we compute the Pearson correlation coefficient $r$ separately for LDA, GGA, and meta-GGA functionals within each panel. In the case of Si, we see a subset of GGA functionals lies below the diagonal, showing relatively large density errors but low density-driven errors, reflected in a moderate $r_{\rm GGA} = 0.75$. This may indicate an error cancellation in the xc energy integral. For NaCl, the correlation is tight and closely follows the $y=x$ line, suggesting that density quality is a strong predictor of density-driven error in the xc energy. This is confirmed by high Pearson coefficients across all rungs ($r_{\rm LDA} = 0.85$, $r_{\rm GGA} = 0.95$, $r_{\rm mGGA} = 0.93$), the latter obtained after excluding the outlier MN12-L. For Cu,  we see a more nuanced picture: while functionals with large density errors tend to show correspondingly large density-driven errors, a significant spread appears near the origin. This region includes several GGA functionals that rank inconsistently between the two metrics, with accurate densities not necessarily corresponding to small density-driven errors. Nonetheless, the Pearson coefficients still reflect a positive correlation across all rungs ($r_{\rm LDA} = 0.68$, $r_{\rm GGA} = 0.89$, $r_{\rm mGGA} = 0.96$).

We also examined the potential correlation between xc functional errors and the dependence of the xc energy density, $\epsilon_\text{xc}(\mathbf{r})$, on the kinetic energy density $\tau(\mathbf{r})$, quantified by the average of $\partial \epsilon_\text{xc}(\mathbf{r})/\partial \tau(\mathbf{r})$~\cite{nazarof2011}. As shown in Figure~\ref{fig:ultra-nonlocality}, no clear systematic trend emerges across the three systems considered: functionals with similar $\tau$-dependence show different errors, and the spread within each system does not follow a monotonic pattern.

\subsection{Practical implications and historical trends}

Understanding whether errors in DFAs stem from the functional form, $\Delta E^F_{\rm xc}$, or from density inaccuracies, $\Delta E^D_{\rm xc}$, has direct practical consequences.  When the functional error dominates, $|\Delta E^F_{\rm xc}| \gg |\Delta E^D_{\rm xc}|$, as observed for most functionals in Cu, improving the functional form itself is the primary path forward. Conversely, when density-driven errors are dominating, as seen with SOGGA11 in Si and NaCl, density correction schemes or hybrid approaches that incorporate higher-quality densities become valuable. This decomposition thus guides toward either functional development or density improvement strategies.  Additionally, our analysis reveals that many functionals benefit from fortuitous error cancellation between $\Delta E^F_{\rm xc}$ and  $\Delta E^D_{\rm xc}$ when used self-consistently. While this may yield accurate results for specific properties in benchmarked systems, it compromises transferability. For instance, 63\% of functionals in Si show $| \Delta E_{\rm xc}| \le |\Delta E^F_{\rm xc}| $, indicating a significant error cancellation, compared to 18\% in Cu. Functionals that minimize both error components independently are more likely to maintain accuracy across diverse systems. This is particularly relevant for DFT-based machine learning potentials, where data generated with functionals with large density-driven errors may encode systematic biases that propagate through the learned potentials, while the material-dependent nature of error cancellation implies that potentials trained on one class of materials may not transfer reliably to others.

\begin{figure}
    \centering
    \includegraphics[width=\linewidth]{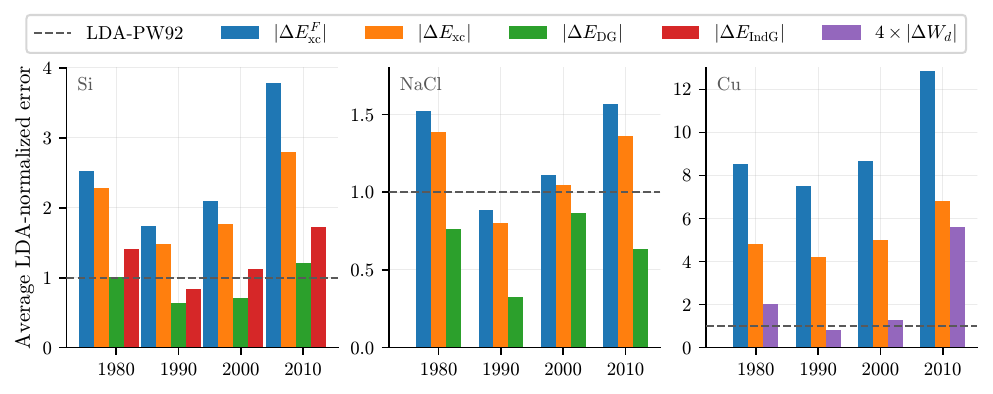}
    \caption{Historical evolution of functional and total xc energy errors, and total errors for direct and indirect KS band gaps and d-band widths, across the studied functionals and systems. The bar plot represents the average absolute LDA-normalized error per decade. Note that $|\Delta W_d|$ is scaled by a factor of 4 for visibility. }
    \label{fig:hist_trend_Ef}
\end{figure}

In Fig.~\ref{fig:hist_trend_Ef}, we examine how functional performance has evolved over the decades by evaluating DFAs on the QMC density for Si, NaCl and Cu. It shows the absolute functional error, $|\Delta E^F_{\rm xc}|$, the total xc energy error, $|\Delta E_{\rm xc}|$, and, where applicable, direct, $|\Delta E_{\rm DG}|$, and indirect, $|\Delta E_{\rm IndG}|$, KS band gap errors, and d-band width errors. All errors are normalized to the corresponding LDA-PW92 values. The results are presented as decade averages to capture the general trends in functional development. For combination of exchange and correlation functionals, the assigned year is the more recent of the publication years of the exchange and correlation components.

Overall, functionals from the 1990s show the best performance across all quantities and all materials. However, on average, they perform worse than LDA-PW92 for both the total xc energy and the functional error in Si and Cu, while showing a clear improvement for NaCl, where they reduce the functional error by about 10\% compared to LDA-PW92. In contrast, modern functionals tend to show larger functional errors, which increase over time and proportionally with the total xc energy error. 

In case of silicon, even with the decreased xc energy accuracy of functionals from the 2000s, they yield an accuracy comparable to the 1990s functionals for the KS band gaps, then the errors show a clear increase for the 2010s decade. 

The trend for sodium chloride shows a deterioration in the quality of the band gap of 2000s functionals compared to the best performers, then the error decreases relatively from 2000s to 2010s. 

For the metal, both the absolute functional error and total xc energy error remain, on average, higher than LDA-PW92. This indicates that ongoing functional development does not translate into better total energy predictions for metals. For the d-band width, the 1990s functionals also provide the most accurate description and along with the 2000s remain better than LDA-PW92 (note the scaling by 4 in the figure). The 2010s decade shows a sharp degradation in performance, with the error exceeding the LDA-PW92.

Our findings for the historical trend, based on the study of functional errors, an intrinsic measure of DFA quality, align with trends observed in total xc energy and density accuracy, as discussed in our previous assessment~\cite{Aouina2024}.

\section{Conclusion}
In summary, across the three materials the percentage of functionals that have 
a total xc error lower than the functional error, $|\Delta E_{\rm xc}| \leq |\Delta E^F_{\rm xc}|$, is 63\% for Si, 28\% for NaCl and 18\% for Cu. 
This trend suggests that in silicon, which is the most benchmarked material in DFT, most functionals benefit from error cancellation when used with their self-consistent density. This error cancellation is less common for the insulator and even less so for the metal, where improving the density leads to a clear improvement of the xc energy for most functionals.

The only functional that maintains a lower functional, $\Delta E^F_{\rm xc}$, error than LDA-PW92 across all three materials is VWN, which systematically overestimates the xc energy. Within the GGA family, PBEP86 demonstrates remarkable robustness, maintaining its position among the top 15 functionals within the GGA family, a distinction not achieved by any other GGA functional. For meta-GGAs we find SCAN and VCML-rVV10. It is interesting to note that these two leading meta-GGAs differ in their design: VCML-rVV10 is  a  multi-purpose empirical functional, while SCAN is completely non-empirical. Despite its empirical nature, VCML-rVV10 still satisfies three of the 17 exact constraints known for semi-local functionals, namely the LDA limit, the exchange gradient expansion, and the cancellation of the spurious Hartree energy in the Hydrogen atom.~\cite{trepte_data_driven_2022}.  

Regarding the systematic improvement of the xc energy across all materials due to the use of accurate QMC density, the functionals that show an improvement of more than 10\% of the total error are PW92, PZ81, and VWN from the LDA family, HCTH407, OPZ81, OVWN, SP86, and SPW91 from the GGA family, and only $\tau$-HCTH as a meta-GGA. For functionals that perform well for both xc energy and KS gaps in Si and NaCl, we typically find combinations of BECKE or GILL exchange with correlation from PBE or its predecessors PW91 or P86. When focusing on the metal and the insulator, PBEsol, PBEOP or PBELYP are the functionals that provide the best xc energy for both systems.   

In both Si and NaCl, SOGGA11 struggles due to its poor density, leading to density-driven errors, $E^D_{\rm xc}$, larger than its intrinsic functional errors. This behavior differs from most other functionals in these systems, where the functional error itself dominates the total error. In contrast to the speculation in Ref.~\onlinecite{mezei_electron_2017}, where the possibility of internal error cancellation in SOGGA11 was discussed without explicit analysis of the functional or density-driven total energy errors, our results show that internal error cancellation does not occur in any of the systems we studied. In another study on water~\cite{dasgupta_elevating_2021, palos_assessing_2022}, it was demonstrated that SCAN also follows this opposite trend where $\Delta E^D_{\rm xc} \gg \Delta E^F_{\rm xc}$. However, for the systems studied here, both SCAN and its revised version, r2SCAN, show the same trend as most other functionals.  

The historical analysis shows that progress in DFA development has been uneven, with functionals from the 1990s generally providing the most balanced performance across different material types. Modern functionals continue to struggle with metallic systems, often performing worse than LDA despite decades of development. The deterioration observed for recent functionals stems largely from meta-GGAs.

Finally, our assessment shows that the best-performing functional depends on the material, with some yielding near-exact xc energies when evaluated on the QMC density. Among the functional families, GGAs are the most transferable, maintaining reasonable accuracy for both xc energies and KS gaps across all three systems. However, we do not observe a systematic improvement in accuracy when ascending Jacob’s ladder~\cite{perdew_jacobs_2001} from the LDA level to GGA and meta-GGA levels; higher-rung functionals do not consistently outperform those on lower rungs across all materials. While our analysis is limited to these three representative materials, the results reveal important limitations of commonly used DFAs, which may have direct consequences for the accuracy of DFT-based machine learning potentials frequently applied in materials simulations.

\section{Data Availability}
The data supporting the findings of this study are available in the published article and the Supporting Information. The full dataset is openly available through the NOMAD Laboratory Archive~\cite{aouina2026nomaddataset}.

\section{Acknowledgments}
We thank Miguel Marques, Lucia Reining and Matteo Gatti for fruitful discussions. A.A. and S.B. gratefully acknowledge funding from the Volkswagen Foundation through the Momentum program (project ``Dandelion").

\bibliography{refs} 
\end{document}


\section{Computational details}

The results presented in this work are based on a systematic analysis of functional and density-driven errors across 70 exchange-correlation (xc) functionals. All calculations were performed using the \texttt{Density-Functional ToolKit (DFTK.jl)}~\cite{DFTKjcon} with \texttt{Julia} bindings to \texttt{Libxc}~\cite{marques2012,lehtola2018}.

\subsection{DFT computational parameters}
To ensure direct compatibility and fair comparison with the quantum Monte Carlo (QMC) benchmarks, we adopted identical calculation parameters to those used in the reference study~\cite{Chen2021}. Specifically:
\begin{itemize}
\item \textbf{Crystal structures and lattice parameters:} Silicon calculations were carried out using a 2-atom primitive cell with a conventional lattice parameter of 10.263087~Bohr. Sodium chloride was treated with the primitive cell of the rocksalt structure at a conventional lattice parameter of 10.7563~Bohr. For copper, a conventional 4-atom cubic cell with a lattice parameter of 6.790~Bohr was adopted.

\item \textbf{Pseudopotentials:} Multiple-projector optimized norm-conserving pseudopotentials\cite{NY_psp} (ONCVPSP), generated following the method of Hamann\cite{Hamann2013}, were used throughout, without non-linear core corrections. For silicon, Ne-shell states were treated as core and the $3s$, $3p$, and $3d$ orbitals as valence. For sodium chloride, a [He] core was assigned to Na and a [Ne] core to Cl. Copper was treated with a [Ne] core.

\item \textbf{Kinetic energy cutoffs:} Plane-wave kinetic energy cutoffs were chosen to match the QMC reference values: 12.5~Ha for silicon, 20~Ha for sodium chloride, and 32~Ha for copper.

\item \textbf{K-point grids:} Brillouin zone sampling was performed on $6 \times 6 \times 6$ Monkhorst-Pack grids for all three materials.
\end{itemize}

\section{List of exchange-Correlation functionals}

Tables~\ref{tab:lda_list}-\ref{tab:mgga_list} provide the complete list of 70 exchange-correlation functionals considered in this work, along with their rung classification, publication year, and corresponding references.


\renewcommand{\thetable}{S\arabic{table}}
\begin{table}[htbp]
\centering
\caption{List of Local Density Approximation (LDA) functionals.}
\label{tab:lda_list}
\footnotesize
\begin{tabular}{cll}
\toprule
\# & Functional & Year \\
\midrule
1 & PW92 \cite{Dirac1930_376, Bloch1929_545, Perdew1992_13244} & 1992 \\
2 & PZ81 \cite{Dirac1930_376, Bloch1929_545, Perdew1981_5048} & 1981 \\
3 & SLATTER \cite{Dirac1930_376, Bloch1929_545} & 1974 \\
4 & VWN \cite{Dirac1930_376, Bloch1929_545, Vosko1980_1200} & 1980 \\
5 & VWN5RPA \cite{Dirac1930_376, Bloch1929_545, Vosko1980_1200} & 1980 \\
6 & XALPHA \cite{Dirac1930_376, Bloch1929_545, Slater1951_385} & 1974 \\
7 & ZLP-LDA \cite{Zhao1993_918} & 1993 \\
\bottomrule
\end{tabular}
\end{table}
\begin{table}[htbp]
\centering
\caption{List of Generalized Gradient Approximation (GGA) functionals (1/2).}
\label{tab:gga_list_1}
\footnotesize
\begin{tabular}{cll}
\toprule
\# & Functional & Year \\
\midrule
1 & BECKE (exchange only) \cite{Becke1988_3098} & 1988 \\
2 & BLYP (BECKE exchange \cite{Becke1988_3098} + LYP correlation \cite{Lee1988_785, Miehlich1989_200}) & 1988 \\
3 & BOP (BECKE exchange \cite{Becke1988_3098} + OP correlation \cite{Tsuneda1999_10664}) & 2001 \\
4 & BP86 (BECKE exchange \cite{Becke1988_3098} + P86 correlation \cite{Perdew1986_8822}) & 1986 \\
5 & BPBE (BECKE exchange \cite{Becke1988_3098} + PBE correlation \cite{Perdew1996_3865, Perdew1997_1396}) & 1996 \\
6 & BPW91 (BECKE exchange \cite{Becke1988_3098} + PW91 correlation \cite{Perdew1991, Perdew1992_6671, Perdew1993_4978}) & 1991 \\
7 & BPZ81 (BECKE exchange \cite{Becke1988_3098} + PZ81 correlation \cite{Perdew1981_5048}) & 1988 \\
8 & BVWN (BECKE exchange \cite{Becke1988_3098} + VWN correlation \cite{Vosko1980_1200}) & 1988 \\
9 & BVWN5RPA (BECKE exchange \cite{Becke1988_3098} + VWN5RPA correlation \cite{Vosko1980_1200}) & 1988 \\
10 & EDF1 \cite{Adamson1998_6} & 1998 \\
11 & GILL (exchange only) \cite{Gill1996_433} & 1996 \\
12 & GLYP (GILL exchange \cite{Gill1996_433} + LYP correlation \cite{Lee1988_785, Miehlich1989_200}) & 1996 \\
13 & GOP (GILL exchange \cite{Gill1996_433} + OP correlation \cite{Tsuneda1999_10664, Tsuneda1999_5656}) & 2001 \\
14 & GP86 (GILL exchange \cite{Gill1996_433} + P86 correlation \cite{Perdew1986_8822}) & 1996 \\
15 & GPBE (GILL exchange \cite{Gill1996_433} + PBE correlation \cite{Perdew1996_3865, Perdew1997_1396}) & 1996 \\
16 & GPW91 (GILL exchange \cite{Gill1996_433} + PW91 correlation \cite{Perdew1991, Perdew1992_6671, Perdew1993_4978}) & 2001 \\
17 & HCTH407 \cite{Boese2001_5497} & 1997 \\
18 & MOHLYP \cite{Schultz2005_11127} & 2005 \\
19 & MPWLYP1W \cite{Dahlke2005_15677} & 2005 \\
20 & N12 \cite{Peverati2012_2310} & 2012 \\
21 & OLYP (OPTX exchange \cite{Handy2001_403} + LYP correlation \cite{Lee1988_785, Miehlich1989_200}) & 2001 \\
22 & OP86 (OPTX exchange \cite{Handy2001_403} + P86 correlation \cite{Perdew1986_8822}) & 1986 \\
23 & OPBE (OPTX exchange \cite{Handy2001_403} + PBE correlation \cite{Perdew1996_3865, Perdew1997_1396}) & 2001 \\
24 & OPTX (exchange only) \cite{Handy2001_403} & 2001 \\
25 & OPW91 (OPTX exchange \cite{Handy2001_403} + PW91 correlation \cite{Perdew1991, Perdew1992_6671, Perdew1993_4978}) & 2001 \\
26 & OPZ81 (OPTX exchange \cite{Handy2001_403} + PZ81 correlation \cite{Perdew1981_5048}) & 2001 \\
27 & OVWN (OPTX exchange \cite{Handy2001_403} + VWN correlation \cite{Vosko1980_1200}) & 2001 \\
\bottomrule
\end{tabular}
\end{table}

\begin{table}[htbp]
\centering
\caption{List of Generalized Gradient Approximation (GGA) functionals (2/2).}
\label{tab:gga_list_2}
\footnotesize
\begin{tabular}{cll}
\toprule
\# & Functional & Year \\
\midrule
28 & OVWN5RPA (OPTX exchange \cite{Handy2001_403} + VWN5RPA correlation \cite{Vosko1980_1200}) & 2001 \\
29 & PBE (PBE exchange \cite{Perdew1996_3865, Perdew1997_1396} + PBE correlation \cite{Perdew1996_3865, Perdew1997_1396}) & 1996 \\
30 & PBELYP (PBE exchange \cite{Perdew1996_3865, Perdew1997_1396} + LYP correlation \cite{Lee1988_785, Miehlich1989_200}) & 1997 \\
31 & PBELYP1W \cite{Dahlke2005_15677} & 2005 \\
32 & PBEOP (PBE exchange \cite{Perdew1996_3865, Perdew1997_1396} + OP correlation \cite{Tsuneda1999_10664, Tsuneda1999_5656}) & 1997 \\
33 & PBEP86 (PBE exchange \cite{Perdew1996_3865, Perdew1997_1396} + P86 correlation \cite{Perdew1986_8822}) & 1996 \\
34 & PBEPW91 (PBE exchange \cite{Perdew1996_3865, Perdew1997_1396} + PW91 correlation \cite{Perdew1991, Perdew1992_6671, Perdew1993_4978}) & 1997 \\
35 & PBEPZ81 (PBE exchange \cite{Perdew1996_3865, Perdew1997_1396} + PZ81 correlation \cite{Perdew1981_5048}) & 1997 \\
36 & PBEVWN (PBE exchange \cite{Perdew1996_3865, Perdew1997_1396} + VWN correlation \cite{Vosko1980_1200}) & 1997 \\
37 & PBEX (exchange only) \cite{Perdew1996_3865, Perdew1997_1396} & 1997 \\
38 & PBEsol (PBEsol exchange \cite{Perdew2008_136406} + PBEsol correlation \cite{Perdew2008_136406}) & 2008 \\
39 & PW91 (PW91 exchange \cite{Perdew1991, Perdew1992_6671, Perdew1993_4978} + PW91 correlation \cite{Perdew1991, Perdew1992_6671, Perdew1993_4978}) & 1992 \\
40 & PW91LYP (PW91 exchange \cite{Perdew1991, Perdew1992_6671, Perdew1993_4978} + LYP correlation \cite{Lee1988_785, Miehlich1989_200}) & 1991 \\
41 & PW91P86 (PW91 exchange \cite{Perdew1991, Perdew1992_6671, Perdew1993_4978} + P86 correlation \cite{Perdew1986_8822}) & 1991 \\
42 & PW91PBE (PW91 exchange \cite{Perdew1991, Perdew1992_6671, Perdew1993_4978} + PBE correlation \cite{Perdew1996_3865, Perdew1997_1396}) & 1997 \\
43 & PW91PZ81 (PW91 exchange \cite{Perdew1991, Perdew1992_6671, Perdew1993_4978} + PZ81 correlation \cite{Perdew1981_5048}) & 1991 \\
44 & PW91VWN (PW91 exchange \cite{Perdew1991, Perdew1992_6671, Perdew1993_4978} + VWN correlation \cite{Vosko1980_1200}) & 1991 \\
45 & PW91X (exchange only) \cite{Perdew1991, Perdew1992_6671, Perdew1993_4978} & 1991 \\
46 & RPBE (RPBE exchange \cite{Hammer1999_7413, Perdew1996_3865, Perdew1997_1396} + PBE correlation \cite{Perdew1996_3865, Perdew1997_1396}) & 1999 \\
47 & SLYP (SLATER exchange \cite{Dirac1930_376, Bloch1929_545} + LYP correlation \cite{Lee1988_785, Miehlich1989_200}) & 1988 \\
48 & SOGGA (SOGGA exchange \cite{Zhao2008_184109} + PBE correlation \cite{Perdew1996_3865, Perdew1997_1396}) & 2008 \\
49 & SOGGA11 \cite{Peverati2011_1991} & 2011 \\
50 & SP86 (SLATER exchange \cite{Dirac1930_376, Bloch1929_545} + P86 correlation \cite{Perdew1986_8822}) & 1986 \\
51 & SPW91 (SLATER exchange \cite{Dirac1930_376, Bloch1929_545} + PW91 correlation \cite{Perdew1991, Perdew1992_6671, Perdew1993_4978}) & 1991 \\
52 & mPWPBE (mPW exchange \cite{Adamo1998_664} + PBE correlation \cite{Perdew1996_3865, Perdew1997_1396}) & 1998 \\
53 & revPBE (revPBE exchange \cite{Zhang1998_890} + PBE correlation \cite{Perdew1996_3865, Perdew1997_1396}) & 1998 \\
\bottomrule
\end{tabular}
\end{table}

\begin{table}[htbp]
\centering
\caption{List of meta-Generalized Gradient Approximation (meta-GGA) functionals.}
\label{tab:mgga_list}
\footnotesize
\begin{tabular}{cll}
\toprule
\# & Functional & Year \\
\midrule
1 & B97M-V \cite{Mardirossian2015_074111} & 2015 \\
2 & MN12-L \cite{Peverati2012_13171} & 2012 \\
3 & MS0 \cite{Sun2012_051101} & 2012 \\
4 & MS1 \cite{Sun2013_044113} & 2013 \\
5 & MS2 \cite{Sun2013_044113} & 2013 \\
6 & MVS \cite{Sun2015_685} & 2015 \\
7 & SCAN \cite{Sun2015_036402} & 2015 \\
8 & VCML-rVV10 \cite{Trepte2022_1104} & 2022 \\
9 & r2SCAN \cite{Furness2020_8208, Furness2020_9248} & 2020 \\
10 & tauHCTH \cite{Boese2002_9559} & 2002 \\
\bottomrule
\end{tabular}
\end{table}

\bibliography{supp-references}